\documentclass[10pt,twocolumn]{article}

\usepackage{times}
\usepackage[margin=1in, columnsep=0.25in]{geometry}
\usepackage[small]{caption}
\usepackage{comment}
\usepackage{microtype}
\usepackage{fancyvrb,wrapfig}

\usepackage{algorithm}
\usepackage{algpseudocode}
\usepackage{amsmath}
\usepackage{appendix}
\usepackage{balance}
\usepackage{caption}
\usepackage{color}
\usepackage{epstopdf}
\usepackage{graphicx}
\usepackage{listings}
\usepackage{multirow}
\usepackage{pdfpages}
\usepackage{subfig}
\usepackage{tabularx}
\usepackage[compact]{titlesec}
\usepackage{url}
\usepackage{xcolor}

\definecolor{dark-gray}{gray}{0.2}

\newenvironment{packed_item}{
\begin{list}{$\bullet$}{
  \setlength{\itemsep}{-2pt}
  \setlength{\parskip}{1pt}
  \setlength{\labelwidth}{15 pt}
  \setlength{\leftmargin}{10pt}
  \setlength{\itemindent}{0pt}}
}{\end{list}}

\newtheorem{defi}{Definition}
\titlespacing\section{0pt}{12pt plus 4pt minus 2pt}{0pt plus 2pt minus 2pt}
\titlespacing\subsection{0pt}{12pt plus 4pt minus 2pt}{0pt plus 2pt minus 2pt}
\titlespacing\subsubsection{0pt}{12pt plus 4pt minus 2pt}{0pt plus 2pt minus 2pt}

\begin{document}

\title{Tupleware: Redefining Modern Analytics}

\author{
\small{Andrew Crotty, Alex Galakatos, Kayhan Dursun, Tim Kraska, Ugur Cetintemel, Stan Zdonik} \\
\small{Department of Computer Science, Brown University} \\
\small{\{crottyan, agg, kayhan, kraskat, ugur, sbz\}@cs.brown.edu}
}
\date{}
\maketitle

\begin{abstract}
There is a fundamental discrepancy between the targeted and actual users of current analytics frameworks.
Most systems are designed for the data and infrastructure of the Googles and Facebooks of the world---petabytes of data distributed across large cloud deployments consisting of thousands of cheap commodity machines.
Yet, the vast majority of users operate clusters ranging from a few to a few dozen nodes, analyze relatively small datasets of up to several terabytes, and perform primarily compute-intensive operations.
Targeting these users fundamentally changes the way we should build analytics systems.

This paper describes the design of Tupleware, a new system specifically aimed at the challenges faced by the typical user.
Tupleware's architecture brings together ideas from the database, compiler, and programming languages communities to create a powerful end-to-end solution for data analysis.
We propose novel techniques that consider the data, computations, and hardware together to achieve maximum performance on a case-by-case basis.
Our experimental evaluation quantifies the impact of our novel techniques and shows orders of magnitude performance improvement over alternative systems.
\end{abstract}

\section{Introduction}
The countless possibilities of advanced analytics have elicited more interest than ever in ``big data" from companies and researchers alike.
Still, current analytics frameworks like Hadoop \cite{hadoop} and Spark \cite{spark} are designed specifically to meet the needs of giant Internet companies; they are built to process petabytes of data in cloud deployments consisting of thousands of cheap commodity machines.
The widespread popularity of advanced analytics, though, has drastically changed the typical use cases.
Nowadays, these frameworks are usually deployed on smaller clusters with more reliable hardware, rather than large cloud deployments.
In fact, it was reported in 2011 that the median Hadoop installation was smaller than 30 nodes \cite{hadoop_blog}.
Furthermore, common MapReduce-style jobs, even at companies as big as Facebook, rarely exceed a few terabytes in size \cite{cluster, cloudera}, making it possible to fit all data in memory on small clusters for these workloads.

Supporting the typical user, then, fundamentally changes the way we should design analytics tools.
Current analytics frameworks are built around the major bottlenecks of large cloud deployments, in which data movement (disk to machine and across the network) is the primary performance bottleneck, machines are slow, and failures are the norm~\cite{mapreduce}.
Conversely, with smaller clusters ranging in size from a few to a few dozen nodes, failures are the exception.
Most importantly, whereas single-node performance is largely irrelevant in cloud deployments, it can no longer be ignored when targeting small clusters.

In this paper we describe Tupleware, a new system designed for typical analytics workloads characterized by relatively small data and compute-intensive operations.
Tupleware compiles workflows comprised of \emph{user-defined functions} (UDFs) directly into a self-contained distributed executable, integrating the LLVM~\cite{llvm} compiler framework to provide a language-agnostic API.
With Tupleware, we (1) address the unique frontend requirements of complex workflows, (2) apply low-level optimizations on a case-by-case basis by considering specific hardware features (e.g., SIMD vectorization, memory bandwidth), and (3) tailor the deployment architecture to more typical hardware configurations.

Our benchmarks, based on common machine learning tasks, demonstrate that our novel techniques achieve orders of magnitude performance improvements over alternative systems like Spark and Hadoop.
In summary, we make the following contributions:
\begin{packed_item}
  \item We present Tupleware, a general analytics system that considers the data, computations, and underlying hardware together in order to fully synthesize a self-contained and highly optimized distributed program.
  \item We propose a new programming model founded in functional programming with monads, allowing for the concise expression of complex workflows while retaining strong optimization potential.
   \item We describe a novel code generation strategy that applies optimizations on a case-by-case basis by examining the internals of UDFs.
   \item We benchmark Tupleware using several common machine learning tasks and show speedups of up to three orders of magnitude over other systems.
\end{packed_item}

\section{System Overview}
\label{sec:system_overview}
\begin{figure}
  \begin{center}
    \includegraphics[width=\columnwidth, trim = 25mm 125mm 25mm 0mm]{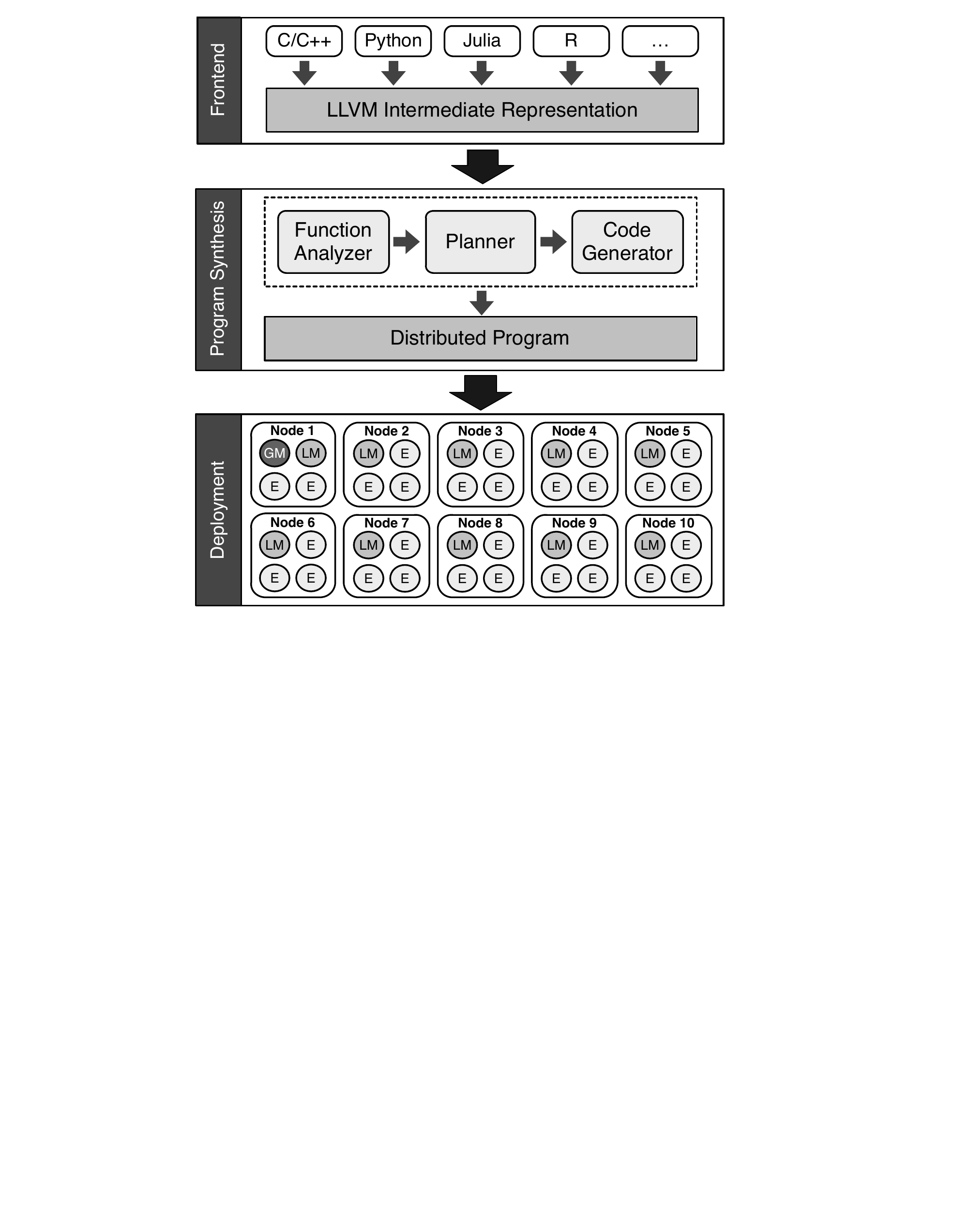}
    \caption{An overview of Tupleware's architecture, which depicts the three distinct yet interrelated components of the system: (1) frontend, (2) program synthesis, and (3) deployment.}
    \label{fig:architecture}
  \end{center}
\end{figure}

Tupleware is a distributed, in-memory analytics platform that targets complex computations such as machine learning (ML) and predictive modeling.
The system architecture is shown in Figure~\ref{fig:architecture} and is comprised of three distinct parts.

\textbf{Frontend:}
Similar to other recent frameworks (e.g., Spark \cite{spark}, Stratosphere \cite{stratosphere}, DryadLINQ \cite{dryadlinq}), Tupleware allows users to define complex workflows directly inside a host language by supplying UDFs to API operators like map and reduce.
Our new algebra, based on the strong foundation of functional programming with monads, seeks a middle ground between flexibility and optimizability while also addressing the unique needs of ML algorithms.
Furthermore, by leveraging the LLVM compiler framework, Tupleware's frontend is language-agnostic, and users can choose from a wide variety of programming languages (visualized as the top boxes in Figure~\ref{fig:architecture}) with little associated overhead.
We describe Tupleware's algebra and API in Section~\ref{sec:frontend}.

\textbf{Program Synthesis:}
When the user submits a job to Tupleware, the \textit{Function Analyzer} examines each UDF to gather statistics for predicting execution behavior.
The \textit{Planner} then translates the workflow to an abstract plan and applies high-level optimizations.
Finally, the \textit{Code Generator} converts the plan into a self-contained distributed program and applies novel optimizations that specifically target the underling hardware using the gathered UDF statistics.
Program synthesis and the accompanying optimizations are explained in Sections~\ref{sec:program_synthesis}-\ref{sec:optimizations}.

\textbf{Deployment:}
After Tupleware has generated the distributed program, the job is automatically scheduled, deployed, and executed on the cluster, depicted in Figure~\ref{fig:architecture} as ten nodes (shown as boxes) each with four hyperthreads (circles inside the boxes).
Tupleware utilizes a multitiered deployment setup, assigning specialized tasks to dedicated threads, and also takes unique approaches to memory management, load balancing, and recovery.
We discuss all of these aspects of deployment further in Section~\ref{sec:deployment}.

\section{Frontend}
\label{sec:frontend}
Ideally, developers want the ability to concisely express complex workflows in their language of choice without having to consider low-level optimizations or the intricacies of distributed execution.
In this section, we describe how Tupleware addresses these points.

\subsection{Background}
The MapReduce \cite{mapreduce} paradigm is a popular programming model for parallel data processing that consists of two primary operators: a \textit{map} that applies a function to every key-value pair, and a \textit{reduce} that aggregates values grouped by key.
Many have criticized MapReduce \cite{dewitt}, in particular for rejecting the advantages of high-level languages like SQL.
However, SQL is unwieldy for expressing many classes of problems, including ML tasks.
For instance, the SQL representation of one ML algorithm described in Section~\ref{sec:evaluation} required four levels of nested subqueries, compared to only a few short map and reduce operations.

Many ML algorithms are most naturally expressed iteratively, but neither MapReduce nor SQL effectively supports iteration.
Unsurprisingly, a number of iterative extensions have been proposed \cite{haloop,twister,sql03}, but users frequently need to rethink algorithmic structure to fit the supplied API, often sacrificing some efficiency in the process.

Furthermore, no existing framework incorporates an elegant and efficient solution for the key ingredient of ML algorithms, namely shared state.
Many attempts to support distributed shared state within a MapReduce-style framework impose substantial restrictions on how and when global values could be used.
For instance, the Map-Reduce-Update model \cite{mrupdate} supplies traditional map and reduce functions with read-only copies of global state values that are recalculated during the update phase after each iteration, but this model is quite restrictive for diverse workflows.
Similarly, Spark provides objects called Accumulators, which are only useful for simple count or sum aggregations on a single key, and their values cannot be accessed from within the workflow.

\subsection{Programming Model}
We therefore need a programming model that strikes a middle ground between the expressiveness of MapReduce and optimizability of SQL while also supporting the unique requirements of ML algorithms.
Tupleware introduces a new algebra based on the foundation of functional programming with monads to address this challenge.
We define this algebra on a data structure called a \emph{TupleSet}, which encapsulates both the data and its associated state.
\begin{defi}[TupleSet]
\label{def:tupleset}
A TupleSet $T$ is a pair $(R,C)$, where $R$ is a relation, which is a set of n-tuples, and $C$ is a Context, which is a dictionary of key-value pairs.
\end{defi}
In Tupleware's algebra, \emph{operators} describe how users can transform a TupleSet:
\begin{defi}[Operator]
\label{def:operator}
An operator $O$ is a second-order function paired with a user defined first-order function $\lambda$ that takes zero or more TupleSets and produces an output TupleSet.
\end{defi}

\begin{figure}
  \begin{center}
    \includegraphics[width=0.75\columnwidth, trim = 25mm 190mm 25mm 10mm]{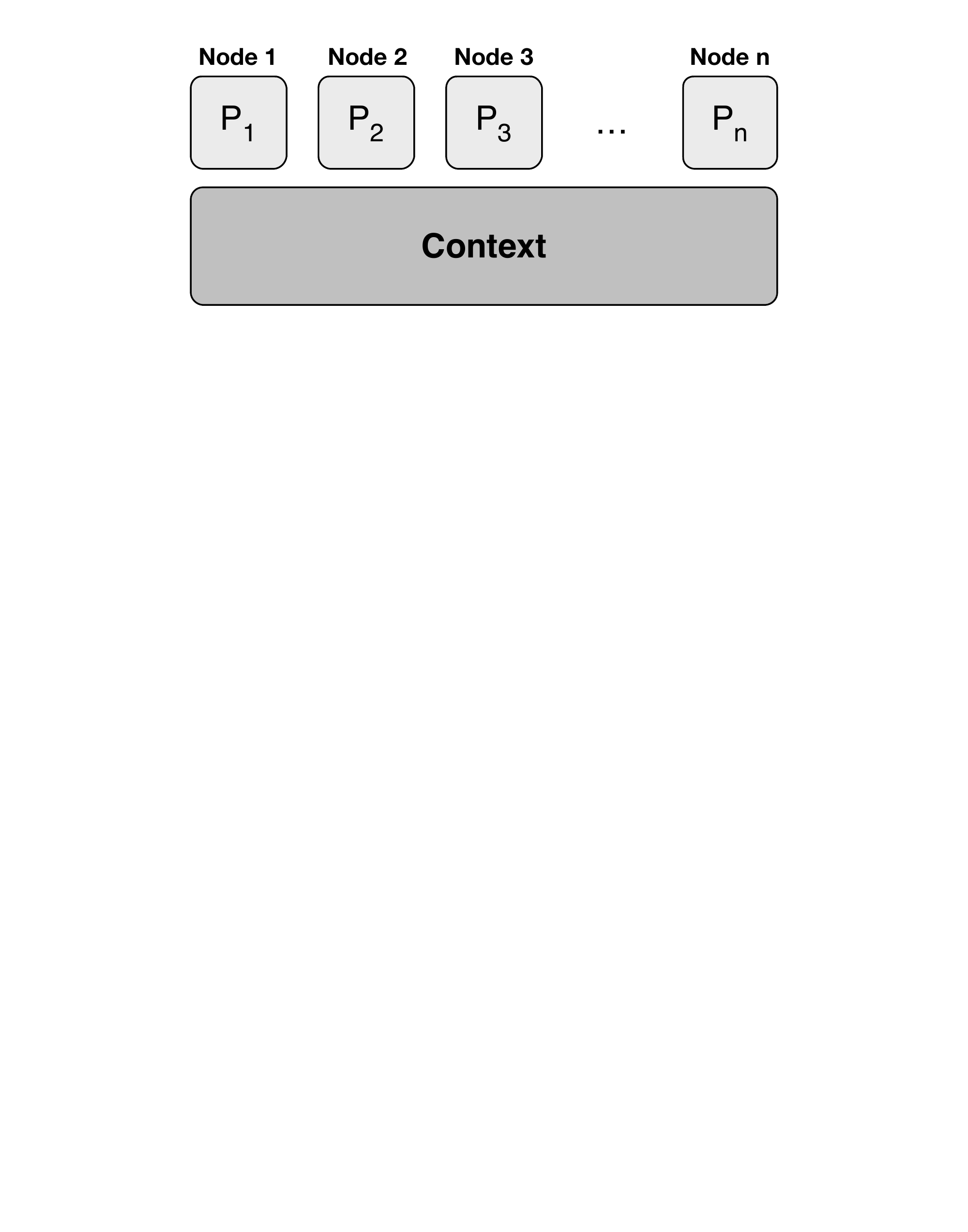}
    \caption{A visualization of Tupleware's logical data model. Partitions $P_1,...,P_n$ of the relation $R$ are spread across $n$ nodes, whereas the Context is logically shared across all nodes.}
    \label{fig:data_model}
  \end{center}
\end{figure}

Tupleware's programming model allows for automatic and efficient parallel data processing.
As shown in Figure~\ref{fig:data_model}, each node in the cluster processes a disjoint subset of the data.
However, unlike other paradigms, Tupleware's API incorporates the notion of global state that is logically shared across all nodes.

\subsection{Operator Types}
We divide operators into four distinct types.
Table~\ref{tab:ops} shows the most common Tupleware operators, as well as the signatures of their associated $\lambda$-functions.
The $\lambda$-functions are supplied by the user and specify the workflow's computation.

\begin{table}[htbp]
  \centering
  \resizebox{\columnwidth}{!}{
  \begin{tabular}{|c|c|c|c|}
    \hline
      \textbf{Class} &
      \textbf{Operator} &
      \textbf{Transformation} &
      \textbf{$\lambda$-Function} \\
    \hline
      \multirow{7}{*}{\small{Relational}} &
      selection$(T)(\lambda)$ &
      $(R, C) \rightarrow (R', C)$ &
      $\sigma_t \rightarrow b$ \\
      &
      projection$(T)(\lambda)$ &
      $(R,C) \rightarrow (R',C)$ &
      $\pi_t \rightarrow t'$ \\
      &
      rename$(T)(\lambda)$ &
      $(R,C) \rightarrow (R',C)$ &
      $\rho_t \rightarrow t'$ \\
      &
      cartesian$(T_1,T_2)$ &
      $(R_1,C_1)(R_2,C_2) \rightarrow (R_1 \times R_2,C_1 \cup C_2)$ &
      - \\
      &
      $\theta$-join$(T_1,T_2)(\lambda)$ &
      $(R_1,C_1)(R_2,C_2) \rightarrow (R',C_1 \cup C_2)$ &
      $\theta_{t_1,t_2} \rightarrow b$ \\
      &
      union$(T_1,T_2)$ &
      $(R_1,C_1)(R_2,C_2) \rightarrow (R_1 \cup R_2,C_1 \cup C_2)$ &
      - \\
      &
      difference$(T_1,T_2)$ &
      $(R_1,C_1)(R_2,C_2) \rightarrow (R_1 \setminus R_2,C_1 \cup C_2)$ &
      - \\
    \hline
      \multirow{3}{*}{\small{Apply}} &
      map$(T)(\lambda)$ &
      $(R,C) \rightarrow (R',C)$ &
      $(t,C) \rightarrow t'$ \\
      &
      flatmap$(T)(\lambda)$ &
      $(R,C) \rightarrow (R',C)$ &
      $(t,C) \rightarrow \{t'\}$ \\
      &
      filter$(T)(\lambda)$ &
      $(R,C) \rightarrow (R',C)$ &
      $(t,C) \rightarrow b$ \\
    \hline
      \multirow{1}{*}{\small{Aggregate}} &
      reduce$(T)(\lambda)(\kappa?)$ &
      $(R,C) \rightarrow (R',C')$ &
      $(t,C) \rightarrow (\Delta_\kappa,C')$ \\
    \hline
      \multirow{5}{*}{\small{Control}} &
      load$()$ &
      $() \rightarrow (R,C)$ &
      - \\
      &
      evaluate$(T)$ &
      $(R,C) \rightarrow (R,C)$ &
      - \\
      &
      save$(T)$ &
      $(R,C) \rightarrow (R,C)$ &
      - \\
      &
      loop$(T)(\lambda)$ &
      $(R,C) \rightarrow (R',C')$ &
      $C \rightarrow b$ \\
      &
      update$(T)(\lambda)$ &
      $(R,C) \rightarrow (R,C')$ &
      $C \rightarrow C'$ \\
    \hline
  \end{tabular}}
  \caption{A subset of TupleSet operators, showing their transformation semantics and $\lambda$-function contracts.}
  \label{tab:ops}
\end{table}

\textbf{Relational:}
Relational operators include all of the traditional SQL transformations.
For example, the user can perform a \textit{selection} by passing a predicate UDF to the corresponding operator.
As given in Table~\ref{tab:ops}, the expected UDF signature has the form: $t \rightarrow b$ where $t \in R$ and $b$ is a Boolean value; that is, the user composes a predicate using the set of operations $\{=,\neq,>,\ge,<,\le\}$ that returns \texttt{true} if a given tuple $t$ of the incoming relation $R$ should be selected for the output relation $R'$ and \texttt{false} otherwise.
Note that relational operators interact only with the relation $R$ of the TupleSet and cannot modify Context variables.
Hence, the Planner described in Section~\ref{sec:program_synthesis:planner} can perform the standard query optimization techniques (e.g., predicate pushdown, join reordering).
Note, though, that operators such as \textit{$\theta$-join} and \textit{union} merge Context variables but do not change their values, performing SQL-style disambiguation of conflicting keys.

\textbf{Apply:}
Apply operators invoke the supplied UDF on every tuple in the relation $R$.
Tupleware's API provides three apply operators: \textit{map}, \textit{flatmap}, and \textit{filter}.
The map operator requires a UDF that specifically produces a \linebreak\textit{1-to-1} mapping (i.e. the UDF takes one input tuple and must return exactly one output tuple).
The flatmap operator takes a UDF that produces a \textit{1-to-N} mapping but is more difficult to optimize.
The filter operator takes a UDF that produces a \textit{1-to-(0:1)} mapping and is less restrictive than the relational selection operator, permitting arbitray predicate logic.
By distinguishing among these different types of apply operators, our programming model provides the system with additional information about the workflow, thereby allowing for greater optimization.

\textbf{Aggregate:}
Aggregate operators perform an aggregation UDF on the relation $R$.
Similar to Spark, Tupleware's \textit{reduce} operator expects a commutative and associative $\lambda$-function.
These semantics allow for the efficient parallelization of computations like sum and count, which return an output relation $R'$ consisting of one or more aggregated values.
Users can also specify a key function $\kappa$ that defines the group-by semantics for the aggregation.
If no key function is provided, then the computation is a single-key reduce (i.e., all tuples have the same key).
Additionally, Tupleware's reduce operator can modify Context variables, as described in Section~\ref{sec:frontend:context}.

\textbf{Control:}
So far, we have not specified how to load data, evaluate a workflow, or save the results.
We assume that the user loads the data into the filesystem as a separate step and then specifies the workflow using the defined operators.
As their names suggest, the \textit{evaluate} and \textit{save} operators actually execute a workflow and store the results, respectively, returning a handle to the result as a new TupleSet that can then be used in a subsequent query.
Notice, though, that this programming model can efficiently cache and reuse results across several computations.
In order to support iterative workflows, which are common to ML algorithms, Tupleware also incorporates a \textit{loop} operator.
The loop operator models a tail recursive execution of the workflow while the supplied loop invariant holds, and the UDF has access to the Context for maintaining information such as iteration counters or convergence criteria.
Finally, Tupleware's algebra provides an \textit{update} operator to allow direct modification of Context variables, which we discuss further in Section~\ref{sec:frontend:context}.

\subsection{Context}
\label{sec:frontend:context}
Tupleware expresses shared state using monads, which are an elegant way to handle side effects in a functional language.
Because of Tupleware's parallel execution model, monads ensure correct concurrent updates to shared state values.
As previously mentioned, this functionality is important for ML algorithms, which can represent models using Context variables.

Changes to the Context as part of a reduce must be commutative and associative.
Conceptually, these updates are not directly applied, but rather added to an update set.
After the operation completes, the deltas stored in the update sets are applied to the context and made visible.
These semantics allow Tupleware to highly parallelize and optimize reduce $\lambda$-functions.

The update operator can directly modify Context variables because it executes logically in a single thread.

\subsection{Language Integration}
As mentioned previously, Tupleware allows users to write workflows and accompanying UDFs in any language with an LLVM compiler, even mixing languages to compose a single job.
Presently, C/C++, Python, Julia, R, and many other languages have LLVM backends.

The system exposes functionality in a given host language via a TupleSet wrapper that implements the Tupleware operator API (see Table~\ref{tab:ops}).
As long as the user adhers to the UDF contracts specified by the API, Tupleware guarantees correct parallel execution.
A TupleSet's Context also has a wrapper that provides special accessor and mutator primitives (e.g., get, increment, decrement).
With the increasing popularity of LLVM, adding new languages is as simple as writing a wrapper to implement Tupleware's API.

\subsection{Example}
\label{sec:frontend:example}
Figure~\ref{fig:python} shows a Python implementation of the \linebreak k-means clustering algorithm using Tupleware's API.
\linebreak K-means is an iterative ML algorithm that classifies each input data item into one of \textit{k} clusters.
In the example, the driver function \texttt{kmeans} defines the workflow using the five specified UDFs, where $t1$ is an input tuple, $t2$ is an output tuple, and $c$ is the Context.
Note that unlike other approaches, Tupleware can store the cluster centroids as Context variables, thereby providing several optimization opportunities (discussed further in Section~\ref{sec:optimizations}).

\begin{figure}
\begin{lstlisting}[basicstyle=\fontsize{5.25}{6.25}\ttfamily,
                     commentstyle=\color{blue},
                     frame=single,
                     language=Python,
                     showstringspaces=false]
ATTR = 2                                 #2 attributes (x,y)
CENT = 3                                 #3 centroids
ITER = 20                                #20 iterations

def kmeans(c):
 ts = TupleSet('data.csv', c)            #load file 'data.csv'
 ts = ts.map(distance)                   #get distance to each centroid
        .map(minimum)                    #find nearest centroid
        .reduce(reassign)                #reassign to nearest centroid
        .update(recompute)               #recompute new centroids
        .loop(iterate)                   #perform 20 iterations
        .evaluate()                      #trigger computation
 return ts.context()['k']                #return new centroids

def distance(t1, t2, c):
 t2.copy(t1, ATTR)                       #copy t1 attributes to t2
 for i in range(CENT):                   #for each centroid:
  t2[ATTR+i] = sqrt(sum(map(lambda       # compute and store distance
     m,n:(n-m)**2,c['k'][i],t1))

def minimum(t1, t2):
 t2.copy(t1, ATTR)                       #copy t1 attributes to t2
 m,n = min(m,n for n,m                   #find index of min distance
   in enumerate(t[:CENT]))
 t2[ATTR] = n                            #assign to nearest centroid

def reassign(t1, c):
 assign = t1[ATTR]                       #get centroid assignment
 for i in range(ATTR):                   #for each attribute:
  c['sum'][assign][i] += t1[i]           # compute sum for assign
 c['ct'][assign] += 1                    #increment count for assign

def recompute(c):
 for i in range(CENT):                   #for each centroid:
  for j in range(ATTR):                  # for each attribute:
   c['k'][i][j] =                        #  calculate average
      c['sum'][i][j]/c['ct'][i]

def iterate(c):
 c['iter'] += 1                          #increment iteration count
 return c['iter'] < ITER                 #check iteration count
\end{lstlisting}
  \caption{A Tupleware implementation of k-means in Python.}
  \label{fig:python}
\end{figure}

\section{Program Synthesis}
\label{sec:program_synthesis}
Once a user has submitted a job, the system (1) examines and records statistics about each UDF, (2) generates an abstract execution plan, and (3) translates the abstract plan into a distributed program.
We refer to this entire process as \emph{program synthesis}.
In this section, we outline the different components that allow Tupleware to synthesize highly efficient distributed programs.

\subsection{Function Analyzer}
\label{sec:program_synthesis:function_analyzer}
Systems that treat UDFs as black boxes have difficulty making informed decisions about how best to execute a given workflow.
By leveraging the LLVM framework, Tupleware can look inside UDFs and determine how to optimize these workflows at a low level.
The Function Analyzer examines the LLVM intermediate representation of each UDF to determine vectorizability, computation cycle estimates, and memory bandwidth predictions.
As an example, Table~\ref{tab:stats} shows the UDF statistics for the k-means algorithm from Section~\ref{sec:frontend:example}.

\begin{table}[htbp]
  \centering
  \resizebox{\columnwidth}{!}{
  \begin{tabular}{|c|c|c|cc|c|}
    \hline
      \multirow{2}{*}{\textbf{Function}} &
      \multirow{2}{*}{\textbf{Type}} &
      \multirow{2}{*}{\textbf{Vectorizable}}  &
      \multicolumn{2}{c|}{\textbf{Compute Time}} &
      \multirow{2}{*}{\textbf{Load Time}} \\
      &
      &
      &
      Predicted &
      Actual &
      \\
    \hline
      \texttt{distance} &
      map &
      yes &
      29 &
      32 &
      3.75 \\
    \hline
      \texttt{minimum} &
      map &
      no &
      17 &
      15 &
      5.62 \\
    \hline
      \texttt{reassign} &
      reduce &
      no &
      15 &
      14 &
      4.22 \\
    \hline
      \texttt{recompute} &
      update &
      no &
      21 &
      23 &
      0 \\
    \hline
  \end{tabular}}
  \caption{Function statistics for the k-means algorithm gathered by the Function Analyzer.}
  \label{tab:stats}
\end{table}

\textbf{Vectorizability:}
Vectorizable UDFs can use \emph{single instruction multiple data} (SIMD) registers to achieve data level parallelism.
For instance, a 256-bit SIMD register on an Intel E5 processor can hold 8$\times$32-bit floating-point values, offering a potential 8$\times$ speedup.
In the k-means example, only the \texttt{distance} UDF is vectorizable, as shown in Table~\ref{tab:stats}.

\textbf{Compute Time:}
One metric for UDF complexity is the number of CPU cycles spent on computation.
CPI measurements \cite{agner} provide cycles per instruction estimates for the given hardware.
Adding together these estimates yields a rough projection for total UDF compute time, but runtime factors (e.g., instruction pipelining, out-of-order execution) can make these values difficult to predict accurately.
However, Table~\ref{tab:stats} shows that these predictions typically differ from the actual measured compute times by only a few cycles.

\textbf{Load Time:}
Load time refers to the number of cycles necessary to fetch UDF operands from memory.
If the memory controller can fetch operands for a particular UDF faster than the CPU can process them, then the UDF is referred to as \emph{compute-bound}; conversely, if the memory controller cannot provide operands fast enough, then the CPU becomes starved and the UDF is referred to as \emph{memory-bound}.
Load time is given by:
\begin{small}
\begin{equation}
  Load \ Time = \frac{Clock \ Speed \times Operand \ Size}{Bandwidth \ per \ Core}
\end{equation}
\end{small}For example, the load time for the \texttt{distance} UDF as shown in Table~\ref{tab:stats} computed on 32-bit floating-point $(x,y)$ pairs using an Intel E5 processor with a 2.8GHz clock speed and 5.97GB/s memory bandwidth per core is calculated as follows:
\begin{small}$3.75 \ cycles = \frac{2.8GHz \times (2 \times 4B)}{5.97GB/s}$\end{small}.

\subsection{Planner}
\label{sec:program_synthesis:planner}
Next, the Planner converts a user's workflow into an abstract plan, which is a logical representation of the job.
Tupleware's frontend supplies the Planner with additional information about the workflow, and the system combines this knowledge with data statistics to apply high-level optimizations (e.g., predicate pushdown, join reordering).
Additionally, the purely functional programming model allows for the integration of other optimizations from the programming lanuage community.
In this regard, we simply apply known techniques, instead focusing on low-level code generation optimizations described in Section~\ref{sec:optimizations}.

\subsection{Code Generator}
\label{sec:program_synthesis:code_generator}
\emph{Code generation} is the process by which compilers translate a high-level language (e.g., Tupleware's algebra) into an optimized low-level form (e.g., LLVM).
As other work has shown \cite{hique}, SQL query compilation techniques can harness the full potential of the underlying hardware, and Tupleware extends these techniques by applying them to the domain of complex analytics.

The Code Generator translates the abstract plan produced by the Planner into a self-contained distributed program and uses UDF statistics gathered by the Function Analyzer to apply low-level optimizations tailored to the underlying hardware.
We discuss these novel optimizations in Section~\ref{sec:optimizations:adaptive}.

As part of the translation process, the Code Generator produces all of the data structure, control flow, synchronization, and communication code necessary to form a complete distributed program.
Unlike other systems that use interpreted execution models, Volcano-style iterators, or remote procedure calls, Tupleware eliminates much associated overhead by compiling in these mechanisms.
Additionally, systems that treat UDFs as black boxes need to rely on external function calls, but Tuplware can inline UDFs to seamlessly merge them with the accompanying support code, providing substantial performance speedups.

\section{Optimizations}
\label{sec:optimizations}
Program synthesis involves a wide range of optimizations that occur on both a logical and physical level.
We divide these optimizations into three categories.

\textbf{DBMS-only:}
As described in Section~\ref{sec:program_synthesis:planner}, Tupleware utilizes well-known query optimization techniques, including predicate pushdown and join reordering.
These DBMS-only optimizations rely on metadata and high-level language semantics, information that is unavailable to compilers.

\textbf{Compiler-only:}
Section~\ref{sec:program_synthesis:code_generator} explains the code generation optimizations that Tupleware leverages, including SIMD vectorization and function inlining.
These compiler-only optimizations occur at a much lower level than DBMSs typically consider.

\textbf{DBMS \& Compiler:}
Some systems incorporate DBMS and compiler optimizations separately, and Sections~\ref{sec:optimizations:pipeline}-\ref{sec:optimizations:operator} describe two such approaches: the (1) \emph{pipeline} and (2) \emph{operator-at-a-time} strategies.
On the other hand, Tupleware combines a high-level algebra and data statistics with the ability to generate code, enabling optimizations that would be impossible for either a DBMS or compiler alone.
Section~\ref{sec:optimizations:adaptive} describes this novel \textit{adaptive strategy}, which applies these hybrid DBMS and compiler optimizations on a case-by-case basis.
To illustrate each of these strategies, we again reference the k-means algorithm from Section~\ref{sec:frontend:example}.

\subsection{Pipeline Strategy}
\label{sec:optimizations:pipeline}
The pipeline strategy \cite{pipeline} aims to maximize data locality by performing as many sequential operations as possible per tuple.
Operations referred to as \emph{pipeline breakers} force the materialization of intermediate results.
For example, a reduce requires an aggregation before advancing to the next phase, whereas all consecutive maps can be pipelined.
Algorithm~\ref{alg:kmeans-pipeline} shows the pipeline approach to k-means.

\begin{algorithm}
  \caption{Pipeline k-means.}
  \begin{algorithmic}
  \begin{scriptsize}
    \State $Data[N]$
    \While{$!converged$}
      \For{$i = 1:N$}
        \State $dist \gets distance(data_i)$
        \State $min \gets minimum(dist)$
        \State $reassign(min)$
      \EndFor
      \State $recompute()$
    \EndWhile
  \end{scriptsize}
  \end{algorithmic}
  \label{alg:kmeans-pipeline}
\end{algorithm}

The pipeline strategy has the major advantage of requiring only a single pass through the data.
Additionally, a tuple is likely to remain in the CPU registers for the duration of processing, resulting in excellent data locality.

\subsection{Operator-at-a-time Strategy}
\label{sec:optimizations:operator}
The operator-at-a-time strategy \cite{monetdb1} performs a single operation at a time for all tuples.
This \emph{bulk processing} approach maximizes instruction locality and opportunities for SIMD vectorization.
Algorithm \ref{alg:kmeans-operator} shows the operator-at-a-time approach to k-means.

\begin{algorithm}
  \caption{Operator-at-a-time k-means.}
  \begin{algorithmic}
  \begin{scriptsize}
    \State $Data[N],Dist[N],Min[N]$
    \While{$!converged$}
      \For{$i = 1:N$}
        \State $dist_i \gets distance(data_i)$
      \EndFor
      \For{$i = 1:N$}
        \State $min_i \gets minimum(dist_i)$
      \EndFor
      \For{$i = 1:N$}
        \State $reassign(min_i)$
      \EndFor
      \State $recompute()$
    \EndWhile
  \end{scriptsize}
  \end{algorithmic}
  \label{alg:kmeans-operator}
\end{algorithm}

The operator-at-a-time strategy, however, requires materialization of intermediate results between each operator, resulting in poor data locality.
A \emph{tiled} variant of this strategy \cite{vectorwise} performs each operation on a cache-resident subset of the data, thus reducing materialization costs and limiting data transfer to the CPU.

\subsection{Adaptive Strategy}
\label{sec:optimizations:adaptive}
In Sections~\ref{sec:optimizations:pipeline} and~\ref{sec:optimizations:operator}, we described two different optimization strategies for code generation.
While each of these approaches has definite advantages in certain situations, they also possess inherent flaws that prevent their universal applicability.
The pipeline strategy boasts excellent data locality but severely limits the advantages of bulk processing, including SIMD vectorization and instruction locality.
On the other hand, the operator-at-a-time strategy benefits greatly from bulk processing but fails to consider data locality, and even the cache-aware tiled variant suffers from the need to materialize intermediate results between operator calls.

Systems that dogmatically adhere to one or the other will necessarily generate suboptimal code in many situations.
A combination of execution strategies, then, is often most sensible, but traditional systems cannot make case-by-case decisions because they regard UDFs as black boxes.
Since Tupleware has the ability to introspect UDFs, we propose a novel \textit{adaptive strategy} that considers the data, computations, and underlying hardware together to generate optimal code in each individual situation.
Furthermore, our strategy can often leverage this knowledge to utilize special data structures when appropriate, offering even greater performance enhancements.
We discuss only optimizations for the map and reduce operators, though we are currently developing optimizations for other operators as well.

\subsubsection{Map}
By default, we group all consecutive maps into a single pipeline to maximize data locality.
Our approach then examines each UDF for SIMD processing opportunities and partitions adjacent maps into vectorizable and nonvectorizable groups.
Intermediate results are materialized between groups in cache-resident blocks.
If the workflow contains no vectorizable UDFs, then the original single-pipeline structure is preserved.

The only exception to this rule arises when a group of one or more vectorizable maps appears at the beginning of a pipeline because of the memory bandwidth bottleneck discussed in Section~\ref{sec:program_synthesis:function_analyzer}.
If the scalar version is already memory-bound, then the group of maps should remain as part of the original pipeline in order to benefit from data locality, since no additional performance increase can be achieved through vectorization.

Consider again the k-means algorithm.
Given the statistics provided by the Function Analyzer in Table~\ref{tab:stats}, we notice that the vectorizable \texttt{distance} UDF is a candidate for pipeline partitioning.
However, since the UDF resides at the beginning of the pipeline, we must also ensure that the computation is not already memory-bound.
In this case, we see that $Compute \ Time > Load \ Time$, so \texttt{distance} is compute-bound and should be split from the pipeline to yield code resembling Algorithm~\ref{alg:kmeans-adaptive}.

\begin{algorithm}
  \caption{Adaptive k-means.}
  \begin{algorithmic}
  \begin{scriptsize}
    \State $Data[N],Dist[N]$
    \While{$!converged$}
      \For{$i = 1:N$}
        \State $dist_i \gets distance(data_i)$
      \EndFor
      \For{$i = 1:N$}
        \State $min \gets minimum(dist_i)$
        \State $reassign(min)$
      \EndFor
      \State $recompute()$
    \EndWhile
  \end{scriptsize}
  \end{algorithmic}
  \label{alg:kmeans-adaptive}
\end{algorithm}

\subsubsection{Reduce}
Generally, a reduce involves maintaining a hash table to store keys and associated aggregates.
Since hash table lookups require random memory accesses, reduce functions cannot be vectorized.
However, the lookup is actually comprised of two distinct parts: the hash calculation and the memory fetch.
The hash calculation can therefore be computed in parallel using SIMD registers, followed by serial execution of memory fetches.

In the case of single-key reduces, we utilize special \emph{reduction variables} to enable the vectorization of aggregation UDFs that are both commutative and associative (e.g., sum).
A reduction variable is transformed into a vector of partial aggregates that are then recombined at the end of the loop to derive the final result.
This optimization completely avoids the usual process of constructing a hash table during the reduce stage.

Additionally, we add reduces to the end of map pipelines to benefit from data locality, as shown by the \texttt{reassign} function in Algorithm \ref{alg:kmeans-adaptive}.

\section{Deployment}
\label{sec:deployment}
After program synthesis (Section~\ref{sec:program_synthesis}), the system now has a self-contained distributed executable.
Each distributed executable contains all necessary communication and synchronization code, avoiding any overhead associated with external function calls.
Tupleware takes a multi-tiered approach to distributed deployment, as shown in Figure~\ref{fig:architecture}.
The system dedicates a single thread on a single node in the cluster as the \textit{Global Manager} (GM), which is responsible for global decisions such as the coarse-grained partitioning of the data among nodes and supervising the current stage of the execution.
In addition, we dedicate one thread per node as a \textit{Local Manager} (LM).
The Local Manager is responsible for the fine-grained management of the local shared memory, as well as for transferring data between machines.
The Local Manager is also responsible for actually deploying compiled programs and does so by spawning new \textit{executor threads} (E), which actually execute the previously compiled program.
During execution, these threads request data from the LM in an asynchronous fashion, and the LM responds with the data and a location for the result.

\subsection{Memory Management}
Similar to DBMSs, Tupleware manages its own memory pool and tries to avoid memory allocations when possible.
Therefore the Local Manager is responsible for keeping track of all active TupleSets and performing garbage collection when necessary.
UDFs that allocate their own memory, though, are not managed by Tupleware's garbage collector.
In addition, we avoid unnecessary object creations or data copying.
For instance, Tupleware often performs updates in-place if the data is not required in subsequent computations.
Additionally, while the Local Manager is idle, it can reorganize and compact the data, as well as free blocks of data that have already been processed.

\subsection{Load Balancing}
Tupleware's data request model is multitiered and pull-based, allowing for automatic load balancing with minimal overhead.
Each Executor thread requests data in small cache-sized blocks from the LM, and each LM in turn requests larger blocks of data from the GM.
All remote data requests occur asynchronously, and blocks are requested in advance to mask transfer latency.

\subsection{Fault Tolerance}
As our experiments demonstrate, Tupleware can process gigabytes of data with sub-second response times, suggesting that checkpointing would do more harm than good.
Extremely long-running jobs on the order of hours or days, though, might benefit from intermediate result recoverability.
In these cases, Tupleware performs simple k-safe checkpoint replication.

However, unlike other systems, Tupleware has a unique advantage: since we fully synthesize distributed programs, we can optionally add these recovery mechanisms on a case-by-case basis.
If our previously described workflow analysis techniques determine that a particular job will have a long runtime, we combine that estimation with the probability of a failure (given our intimate knowledge of the underlying hardware) to decide whether to include recovery code.

\begin{figure*}
\begin{minipage}{10cm}
  \includegraphics[height=3.6cm]{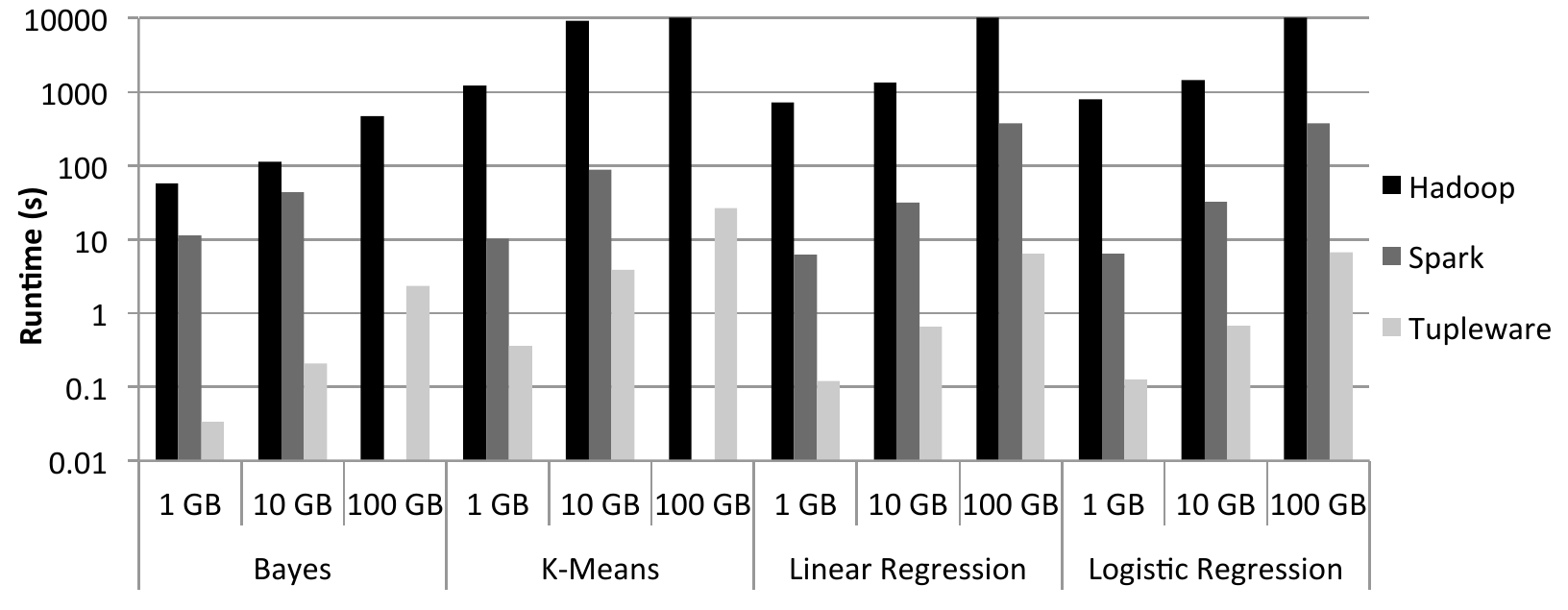}
  \captionof{figure}{8-Node Cluster Performance}
  \label{fig:distributed}
\end{minipage}
\begin{minipage}{7cm}
\centering
  \includegraphics[height=3.2cm, trim=15mm 0mm 0mm 0mm]{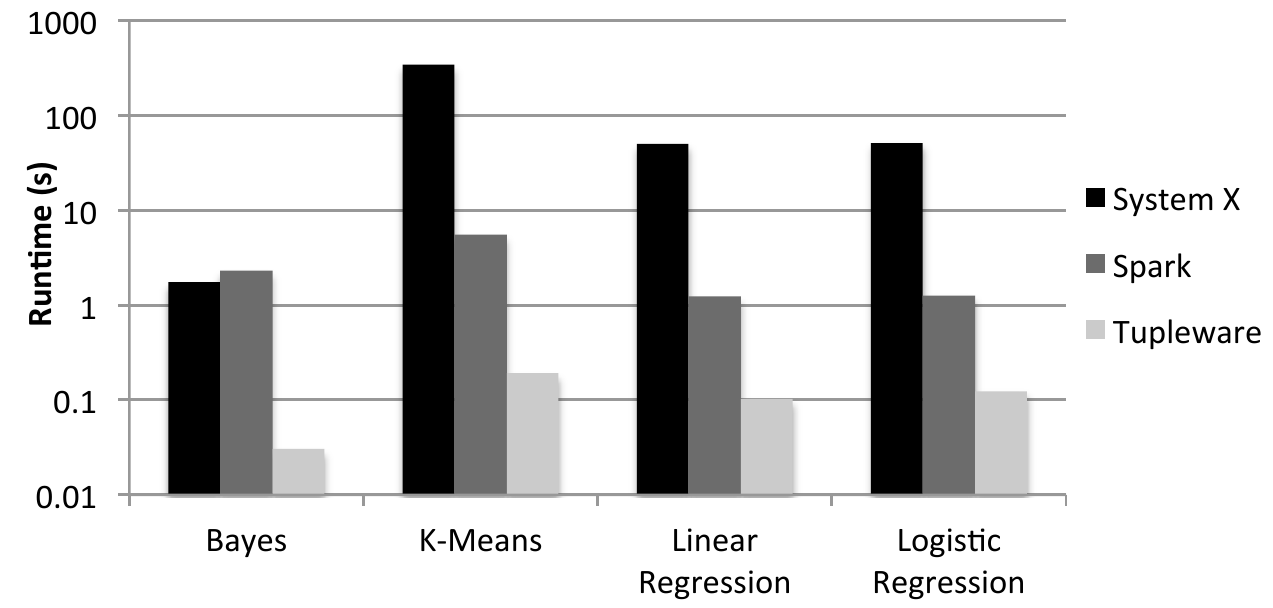}
  \captionof{figure}{Single Node Performance}
  \label{fig:single}
\end{minipage}
\end{figure*}

\begin{figure*}
  \centering
  \begin{scriptsize}
  \begin{tabular}{|c|ccc|ccc|c|c|}
    \hline
      \multirow{3}{*}{\textbf{Algorithm}} &
      \multicolumn{6}{c|}{\textbf{Distributed}} &
      \multicolumn{2}{c|}{\textbf{Single Machine}} \\
      &
      \multicolumn{3}{c|}{\textbf{Hadoop}} &
      \multicolumn{3}{c|}{\textbf{Spark}} &
      \textbf{System X} &
      \textbf{Spark} \\
      &
      1GB &
      10GB &
      100GB &
      1GB &
      10GB &
      100GB &
      &
      \\
    \hline
      K-means &
      $3298\times$ &
      $2379\times$ &
      TO &
      $29\times$ &
      $23\times$ &
      FAIL &
      $1790\times$ &
      $29\times$ \\
    \hline
      Logistic Regression &
      $6334\times$ &
      $2181\times$ &
      TO &
      $51\times$ &
      $49\times$ &
      $55\times$ &
      $416\times$ &
      $10\times$ \\
    \hline
      Linear Regression &
      $5534\times$ &
      $2071\times$ &
      TO &
      $48\times$ &
      $48\times$ &
      $58\times$ &
      $486\times$ &
      $12\times$ \\
    \hline
      Naive Bayes &
      $1709\times$ &
      $544\times$ &
      $203\times$ &
      $341\times$ &
      $210\times$ &
      FAIL &
      $57\times$ &
      $76\times$ \\
    \hline
  \end{tabular}
  \end{scriptsize}
  \caption{Tupleware speedup over existing systems (TO=Timed-Out, FAIL=Memory Failure)}
  \label{tab:speedup}
\end{figure*}

\section{Evaluation}
\label{sec:evaluation}
Our evaluation compares Tupleware's performance to alternative systems in Section~\ref{sec:evaluation:system_benchmarks}, provides a performance breakdown in Section~\ref{sec:evaluation:performance_breakdown}, analyzes the effects of our optimizations in Section~\ref{sec:evaluation:microbenchmarks}, and demonstrates the system's scalability in Section~\ref{sec:evaluation:scalability_benchmarks}.

\subsection{System Benchmarks}
\label{sec:evaluation:system_benchmarks}
To assess the overall performance of our system, we evaluate Tupleware against three widely-used analytics platforms (Hadoop, Spark, and a commercial column DBMS System X) using four common ML tasks.

\subsubsection{Setup}
We benchmark Tupleware in both a distributed cluster and single-machine environment to demonstrate its versatility.
For the distributed case, we compare Tupleware against Spark and Hadoop.
For the single-machine benchmarks, we compare it against Spark, as well as a commercial column DBMS System X.
To execute iterative algorithms on System X, we iteratively call a SQL stored procedure using a JDBC driver program and perform as much of the computation as possible inside the DBMS.
Unless stated otherwise, we used \texttt{c3.8xlarge} instances on Amazon EC2.
These instances have Intel E5-2680v2 processors (10 cores, 25MB Cache), 60 GB RAM, $2\times320$ SSDs, and are connected with 10 Gigabit*4 Ethernet.
The distributed setup consists of $8\times$\texttt{c3.8xlarge} instances.

In all cases except Hadoop, we record the total runtime of each algorithm after the input data has been loaded into memory and parsed, with the caches warmed up.
For all iterative algorithms, we report the total time to complete 20 iterations of the algorithm.

\subsubsection{Workload and Data}
For all systems, we implemented a consistent version of each ML task using the same algorithm with a fixed number of iterations.
We used a combination of real and synthetic datasets to test across a wide range of data characteristics (e.g., size, dimensionality, skew).
For all distributed benchmarks, we generated datasets of 1, 10, and 100GB in size.
We benchmarked four ML algorithms.

\textbf{K-means:}
As described in Section~\ref{sec:frontend:example}, k-means is an iterative clustering algorithm that partitions a dataset into \textit{k} clusters.
To evaluate the performance of each system in a distributed setup, we generated a dataset from three distinct means and ran the algorithm on each system.
To analyze single node performance, we ran k-means using a 70MB synthetic dataset.

\textbf{Logistic Regression:}
Logistic regression attempts to find a hyperplane $w$ that best separates two classes of data by iteratively computing the gradient and updating the parameters of $w$.
We implemented logistic regression with gradient descent on generated data with 1024 features.
We also ran the algorithm using a 10MB subset of the Million Song Dataset \cite{millionsong} in the single node setup.
The dataset contains 26,330 data elements, each with 90 dimensions denoting various audio features.
The algorithm tries to predict whether a given song was produced in 2006 or 2007 based upon these features.

\textbf{Linear Regression:}
Linear regression produces a model by fitting a linear equation to a set of observed data points.
The distributed experiment builds the model by iteratively computing a gradient and updating the weights for a synthetic dataset.
We also ran the algorithm on a single node using the same 10MB subset of the Million Song Dataset.

\textbf{Naive Bayes:}
A naive Bayes classifier is a conditional model that uses feature independence assumptions to assign class labels.
Naive Bayes classifiers are extremely popular and are used for a wide variety of tasks, such as spam filtering, text classification, and sentiment analysis.
For the distributed setup, we trained a naive Bayes classifier on a generated dataset with 128 features and 10 possible labels.
Using a larger subset of the Million Song Dataset, we ran the same naive Bayes algorithm on a single machine in order to predict the release year of a given song.
The 100MB subset contains 90 audio features for 288,070 songs, and we preprocessed the data by mapping continuous values to categorical bins.

\subsubsection{Discussion}
We now discuss the performance of Tupleware against Spark and Hadoop on a small cluster of 8 nodes, and against Spark and System X on a single node.
The results are included in Figures~\ref{fig:distributed}, \ref{fig:single}, and \ref{tab:speedup}.
Overall, our benchmarks show speedups between $203-6334\times$ over Hadoop, $10-341\times$ over Spark, and $57-1790\times$ over System X.

Hadoop encounters substantial I/O overhead when materializing intermediate results to disk between iterations.
On the other hand, Tupleware intelligently caches intermediate results in memory and focuses on compute-bound workloads.
Furthermore, Hadoop's API is not intended for complex analytics and makes optimization difficult.

Spark improves upon Hadoop by storing the working set in memory and offering a richer API.
We therefore measure the greatest speedups for Spark over Hadoop with iterative tasks, whereas the non-iterative naive Bayes runtimes for these two systems are much more similar.
Tupleware can achieve an additional 1 to 2 orders of magnitude improvement over Spark by optimizing at a low level for the computation bottleneck.

System X lacks native support for many fundamentals of ML algorithms, in particular iteration.
Hence, the SQL implementations incurred major performance penalties for k-means and the regressions but not for naive Bayes.

\subsection{Performance Breakdown}
\label{sec:evaluation:performance_breakdown}
No single optimization explains the entire performance gap between Tupleware and other systems.
Only the combination of the frontend, program synthesis, and deployment techniques allows Tupleware to achieve speedups of several orders of magnitude.
In order to explain these speedups, we provide a performance breakdown to quantify the benefits of Tupleware's distinguishing features.
Each experiment isolates one particular feature to measure the idealized contribution to the overall speedup.
This is currently a preliminary breakdown, and we are still investigating other factors.

\begin{figure}
  \centering
  \includegraphics[width=\columnwidth, trim = 0mm 35mm 0mm 25mm]{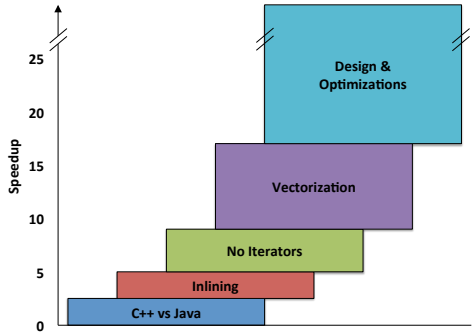}
  \caption{A performance breakdown with idealized runtime speedups achieved by Tupleware over Spark.}
  \label{fig:breakdown}
\end{figure}

\textbf{C++ vs Java:}
We compared the average runtime for a single iteration of each ML algorithm from Section~\ref{sec:evaluation:system_benchmarks} implemented in both C++ and Java.
While C++ offers significantly lower level control, the choice of C++ over Java has relatively little overall impact.
Our experiments achieved a speedup of up to 2.5$\times$.

\textbf{Inlining:}
Compiling queries into executables allows Tupleware to inline functions instead of performing external function calls.
This technique eliminates significant overhead associated with UDF-centric workloads, offering around a 2.5$\times$ speedup.

\textbf{No Iterators:}
Volcano-style iterators recursively call a \texttt{next} function for each operator in the workflow.
This pattern is easily generalized and simple to implement, hence its extensive adoption by a wide range of systems, but has substantial performance penalties.
By dynamically generating code, Tupleware eliminates all overhead associated with iterator-based execution, which explains a factor of up to 4$\times$ in our experiments.

\textbf{Vectorization:}
As mentioned in Section~\ref{sec:program_synthesis:function_analyzer}, SIMD vectorization offers significant performance benefits, and Tupleware can automatically take advantage of this hardware feature.
On the tested hardware, we can see up to an 8$\times$ improvement compared to the scalar version.
Even greater speedups are possible with currently available 512- or 1024-bit SIMD registers.

\textbf{Design \& Optimizations:}
The remainder of the speedup can be attributed to a combination of intangible design differences (e.g., workflow compilation, deployment architecture) and novel optimization techniques.
In Section~\ref{sec:evaluation:microbenchmarks}, we examine the impact of vectorization opportunities made possible by our code generation techniques, as well as our other optimizations.

\subsection{Microbenchmarks}
\label{sec:evaluation:microbenchmarks}
In order to get a better understanding of our code generation techniques, we performed a series of microbenchmarks for (1) map optimizations, (2) single-key reduce optimizations, and (3) Context variables in aggregate UDFs.

\begin{figure*}
  \centering
  \subfloat[][Map strategies]{
  \includegraphics[height=2.7cm]{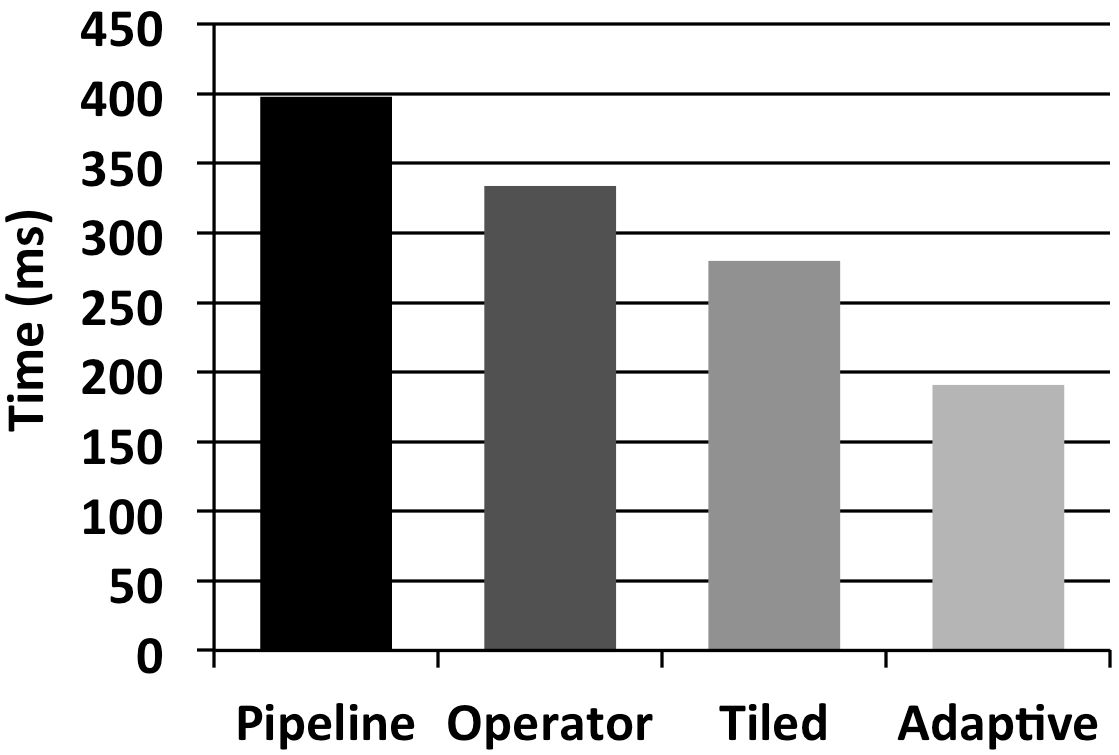}
  \label{fig:map_opt}}
  \subfloat[][Reduction variable]{
  \includegraphics[height=2.7cm]{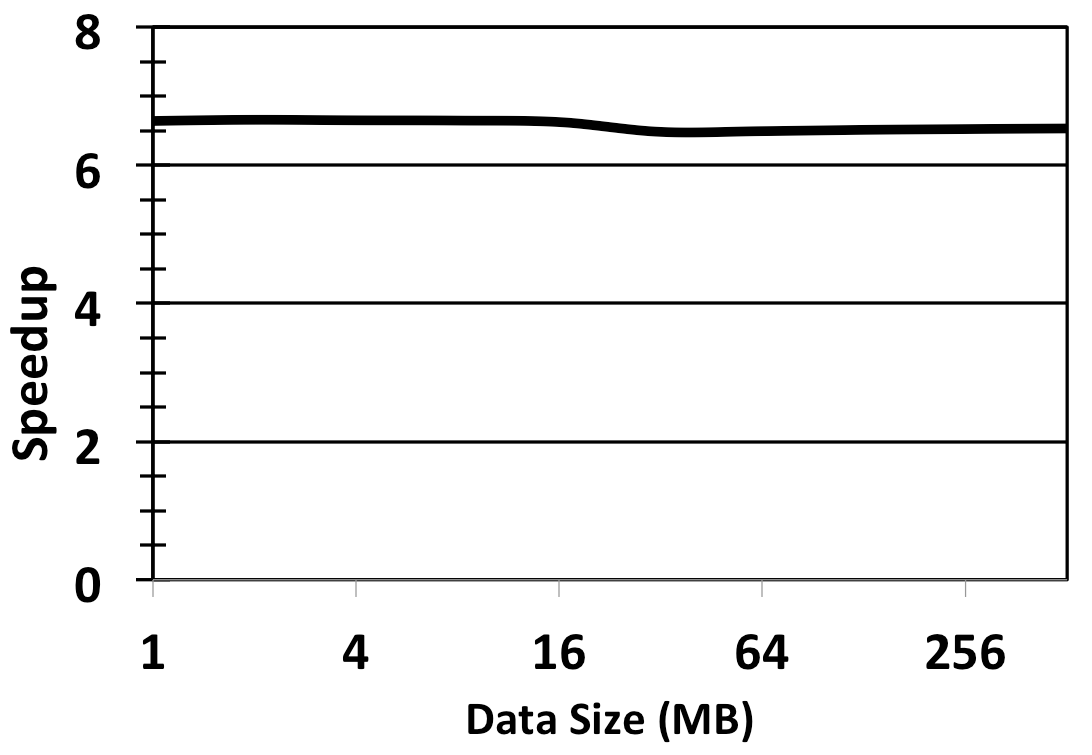}
  \label{fig:red_var}}
  \subfloat[][Context]{
  \includegraphics[height=2.7cm]{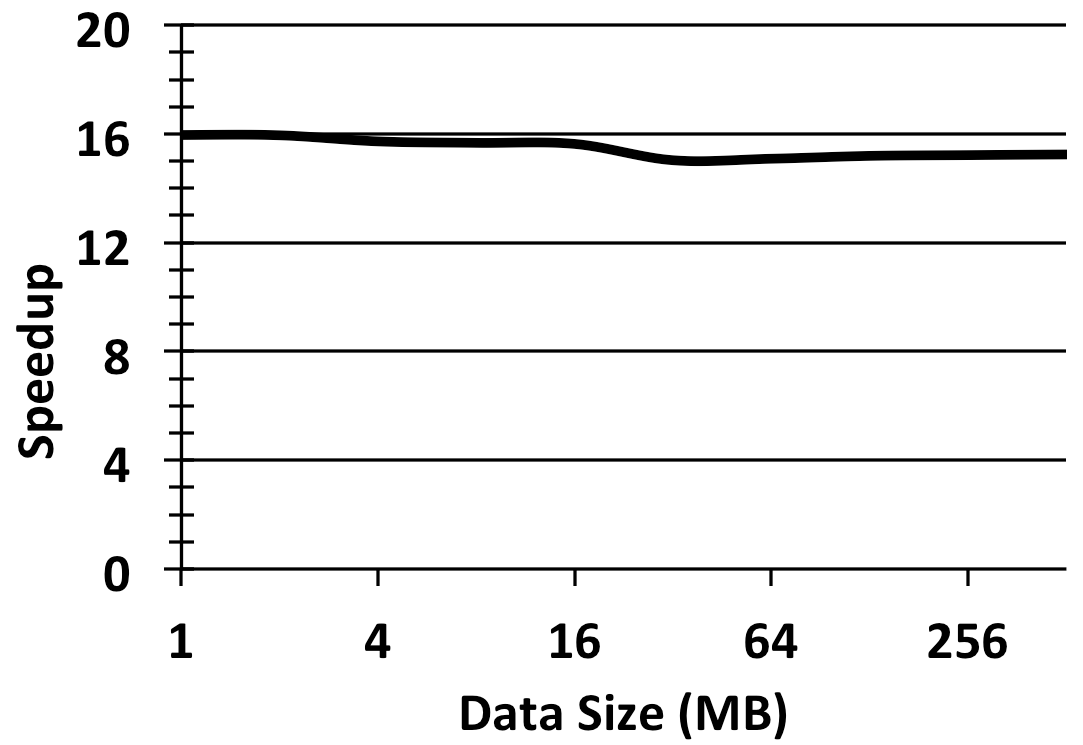}
  \label{fig:context}}
  \subfloat[][Scalability]{
  \includegraphics[height=2.7cm]{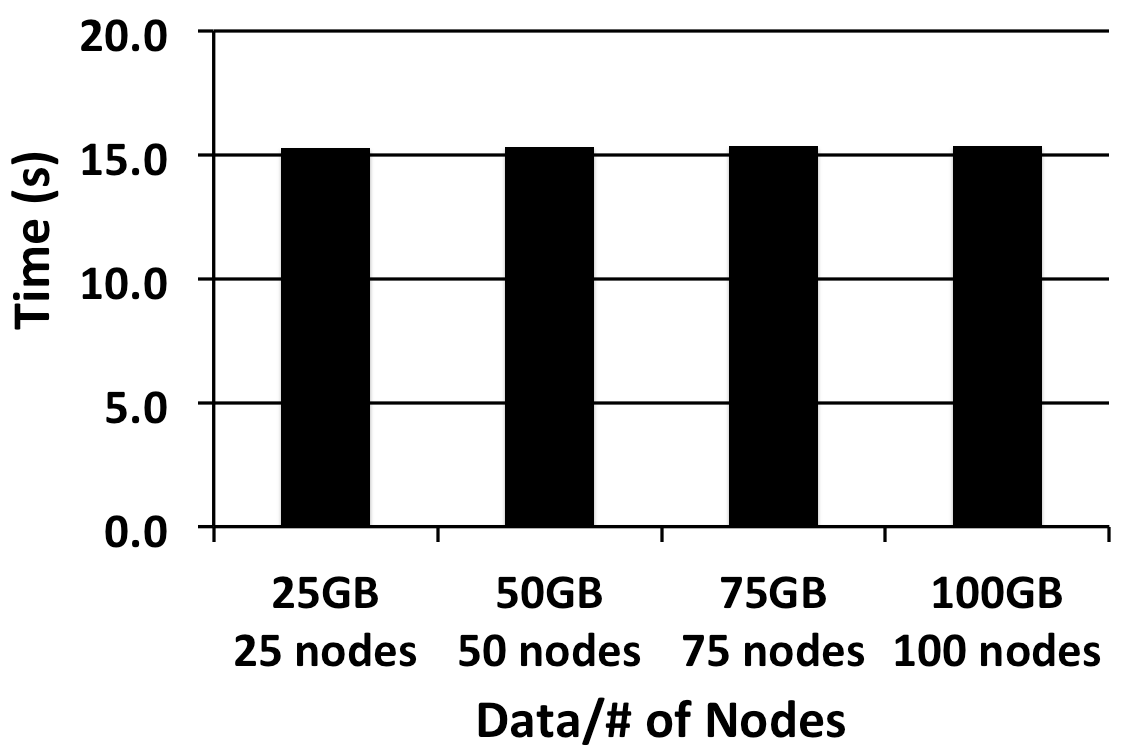}
  \label{fig:scale}}
  \caption{Tupleware microbenchmarks (a-c) and scalability benchmarks (d).}
\end{figure*}

\textbf{Map Strategies:}
We compare our adaptive map optimization strategy to the \emph{pipeline}, \emph{operator-at-a-time}, and \emph{tiled operator-at-a-time} strategies described in Sections~\ref{sec:optimizations:pipeline}-\ref{sec:optimizations:operator}.
We implemented each strategy for k-means in C++, compiled with Clang~3.4, and ran 20 iterations using 70MB of input data on a single \texttt{c3.8xlarge} instance.
The pipeline strategy benefits from excellent data locality by performing all operations consecutively for each data element.
However, this approach prohibits any vectorization due to the fact that the \texttt{minimum} function cannot be vectorized.
The operator-at-a-time strategy applies a single operator once to each data element, materializing results between operators.
Using this approach, the \texttt{distance} function can be vectorized, but materializing intermediate results between subsequent operators incurs significant overhead.
As demonstrated in Figure~\ref{fig:map_opt}, our adaptive strategy performs better than the existing strategies due to the fact that it can take advantage of SIMD vectorization while also pipelining nonvectorizable operations for better data locality.

\textbf{Reduction Variable:}
Based on knowledge about the entire workflow, Tupleware is able to generate code that uses a reduction variable rather than a hash table when performing a single-key aggregation, as described in Section~\ref{sec:optimizations:adaptive}.
Our programming model allows for this optimization and permits SIMD vectorization for single-key aggregation UDFs.
We performed a microbenchmark for a simple reduce with and without a reduction variable.
Figure~\ref{fig:red_var} shows the speedup of the reduction variable over the naive approach for a simple single-key reduce.
This code generation technique yields a consistent speedup of around $6.5\times$ for various data sizes, with a slight drop around 20MB because of the CPU cache size.

\textbf{Context:}
Tupleware is able to highly optimize workflows that use Context variables due to the fact that the system knows about the types and sizes of all Context variables when the program is compiled.
For updates to the Context, Tupleware is able to use direct indexing instead of hashing to perform aggregation.
For example, instead of looking up and incrementing a value for a key in a hash table, Tupleware can directly index into an array and increment the corresponding value.
This eliminates the need to perform a hash function for each key as well as search for the correct key once the bucket is found.
Tupleware is also able to allocate fixed size containers for these Context variables, allowing the system to better utilize the CPU cache.
To measure the impact of this optimization, we compare using hashing to direct indexing for storing a count of 10 distinct keys for datasets of varying size.
As shown in Figure~\ref{fig:context}, this technique improves performance by around $16\times$.

\subsection{Scalability Benchmarks}
\label{sec:evaluation:scalability_benchmarks}
To show how Tupleware performs in clusters of various sizes, we run a \textit{weak scalability} experiment where each machine processes a constant amount of data.
In this experiment, we measure the total runtime of the k-means algorithm on 25, 50, 75, and 100 node clusters, in which each machine processes 1GB of input data.
Each node in the cluster is a \texttt{m3.large} instance with 2 virtual CPUs and 7.5GB of RAM.
Figure~\ref{fig:scale} shows that Tupleware is able to adapt to different cluster sizes and scale linearly to larger clusters.

\section{Related Work}
\label{sec:related_work}
Tupleware's unique design allows the system to highly optimize complex analytics tasks.
While other systems have looked at individual components, Tupleware collectively addresses how to (1) easily and concisely express complex analytics workflows, (2) synthesize self-contained distributed programs optimized at the hardware level, and (3) deploy tasks efficiently on various cluster sizes.

\subsection{Programming Model}
Numerous extensions have been proposed to support iteration and shared state within MapReduce \cite{asterix,haloop,twister}, and some projects (e.g., SystemML~\cite{systemml}) go a step further by providing a high-level language that is translated to MapReduce tasks.
Conversely, Tupleware natively integrates iterations and shared state to support this functionality without sacrificing low-level optimization potential.
Other programming models, such as FlumeJava \cite{flumejava}, Ciel \cite{ciel}, and Piccolo \cite{piccolo} lack the low-level optimization potential that Tupleware's algebra provides.

DryadLINQ \cite{dryadlinq} is most similar to Tupleware's frontend, as it allows the user to incorporate relational transformations into any .NET host language.
The biggest difference between this framework and Tupleware is that DryadLINQ cannot easily express updates to shared state and requires a driver-program for iterations, which precludes many optimizations.

Tupleware also has commonalities with the programming models proposed by Spark~\cite{spark} and Stratosphere~\cite{stratosphere}.
These systems have taken steps in the right direction by providing richer APIs that supply the optimizer with additional information about the workflow, thus permitting standard high-level optimizations.
In addition to these more traditional optimizations, Tupleware's algebra is designed specifically to enable hardware-level optimizations and efficiently handle distributed shared state.

\subsection{Code Generation}
Code generation for query evaluation was proposed as early as System R \cite{systemr}, but this technique has recently gained popularity as a means to improve query performance for in-memory DBMSs \cite{jvm,hique}.
Both HyPer~\cite{hyper1} and VectorWise~\cite{vectorwise} propose different optimization strategies for query compilation, but these systems focus on SQL and do not optimize for UDFs.
In fact, our experiments in Section~\ref{sec:evaluation:microbenchmarks} demonstrate that neither of these strategies is optimal for complex analytics workflows.

LegoBase~\cite{legobase} incorporates a query engine written in Scala that generates specialized C code, allowing for continuous optimization.
Again, LegoBase concentrates on SQL and does not consider complex analytics or UDFs.

Scope~\cite{scope} compiles workflows using the .NET framework.
However, Scope primarily focuses on SQL-like queries against massive datasets rather than complex in-memory analytics and applies no hardware-level optimizations.
Similarly, Tenzing~\cite{tenzing} and Impala~\cite{impala} are SQL compilation engines that also target simple queries on large datasets.

OptiML~\cite{optiml} offers a Scala-embedded, domain-specific language used to generate execution code that targets specialized hardware (e.g., GPUs) on a single machine.
Tupleware on the other hand provides a general, language-agnostic frontend used to synthesize LLVM-based distributed executables for deployment in a cluster.

One key advantage of Tupleware is its ability to adapt to the features of the underlying hardware, including SIMD registers.
Significant work has been done on leveraging SIMD instructions for query processing in traditional DBMSs~\cite{simd1,simd2,simd3}, but the focus is on relational operations and does not consider UDFs.

\subsection{Distribution Architecture}
Hadoop~\cite{hadoop} targets datasets many times larger than the available memory, while Tupleware focuses instead on in-memory processing for complex computations.
Like Tupleware, Spark and Stratosphere also focus on distributed in-memory analytics, but their deployment architecture and iterator-based data processing model impose considerable overhead, as shown in Section~\ref{sec:evaluation:performance_breakdown}.

BID Data Suite~\cite{bid} and Phoenix~\cite{phoenix} are high performance single-node frameworks targeting general analytics, but these systems cannot scale to multiple machines or beyond small datasets.
Scientific computing languages like R~\cite{r} and Matlab~\cite{matlab} have these same limitations.
More specialized systems (e.g., Hogwild!~\cite{hogwild}) provide highly optimized implementations for specific algorithms on a single machine, whereas Tupleware is intended for general computations in a distributed environment.

\section{Conclusion}
\label{sec:conclusion}
Advanced analytics workloads have become commonplace for a wide variety of users.
However, instead of targeting the hardware to which most of these users have access, existing analytics frameworks are designed exclusively for large cloud deployments with thousands of commodity machines.
This paper described the design of Tupleware, a new analytics system geared towards the needs of typical users.  Tupleware combines ideas from the database, compiler, and programming language communities to create a user-friendly yet highly efficient end-to-end solution for complex analytics.
Our experiments demonstrated that Tupleware achieves remarkable speedups of $10-6000\times$ over alternative systems.

\balance
\begin{small}
\bibliographystyle{abbrv}
\bibliography{bib}
\end{small}
\end{document}